\documentstyle{article}
\def\beq{\begin{equation}}
\def\eeq{\end{equation}}
\def\bea{\begin{eqnarray}}
\def\eea{\end{eqnarray}}
\def\nn{\nonumber}
\def\ba{\begin{array}}
\def\ea{\end{array}}   
\def\i{{\rm i}}
\begin{document}
\begin{flushright}
physics/9904064
\end{flushright}

\begin{center}

{\large\bf Quantum mechanical aspects of the halo puzzle}
\footnote{In Proceedings of the 1999 Particle Accelerator Conference~(PAC99)
29 March -- 02 April 1999, New York City,
Editors: A.~Luccio and W.~MacKay}

\bigskip

Sameen Ahmed KHAN and Modesto PUSTERLA \\

Dipartimento di Fisica Galileo Galilei  
Universit\`{a} di Padova \\
Istituto Nazionale di Fisica Nucleare~(INFN) Sezione di Padova \\
Via Marzolo 8 Padova 35131 ITALY \\
E-mail: khan@pd.infn.it, ~~~ http://www.pd.infn.it/$\sim$khan/ \\
E-mail: pusterla@pd.infn.it, ~~~ http://www.pd.infn.it/$\sim$khan/

\end{center}

\bigskip

\begin{abstract}
An interpretation of the the ``halo puzzle'' in accelerators based on
quantum-like diffraction is given. Comparison between this approach 
and the others based on classical mechanics equations is exhibited.
\end{abstract}


\newpage

In this note we point out that, after linearizing the Schr\"{o}dinger-like
equation, for beams in an accelerator one can use the whole apparatus
of quantum mechanics, with a new interpretation of the basic parameters
(for instance the Planck's constant $\hbar \longrightarrow \epsilon$ where 
$\epsilon$ is the normalized beam emittance) and introduce the propagator
$K \left( x_f , t_f | x_i , t_i \right)$ of the
Feynman theory for both longitudinal and transversal motion. 
A procedure of this sort seems particularly effective for a global
description of several phenomena such as intrabeam scattering, space-charge,
particle focusing, that cannot be treated easily in detail by ``classical 
mechanics'' and are considered to be the main cause of the creation of the
``Halo'' around the beam line with consequent losses of particles.

Let us indeed consider the Schr\"{o}dinger like equation for the beam
wave function
\bea
\i \epsilon \partial _t \psi 
= - \frac{\epsilon^2}{2 m} \partial_x ^2 \psi + U \left( x , t \right) \psi
\label{schroedinger-like}
\eea
in the linearized case $U \left (x , t \right)$ does not depend on the 
density $\left| \psi \right|^2$. $\epsilon$ here is the normalized
transversal beam emittance defined as follows:
\bea
\epsilon = m_0 c \gamma \beta \tilde{\epsilon}\,,
\label{epsilon}
\eea
$\tilde{\epsilon}$ being the emittance usually considered, where as
we may introduce the analog of the De Broglie wavelength as
$\lambda = {\epsilon}/{p}$. We now focus our attention on the one 
dimensional transversal motion along the $x$-axis of the beam particles 
belonging to a single bunch and assume a Gaussian transversal profile 
for a particles injected in to a circular machine. We describe all the 
interactions mentioned above, that cannot be treated in detail, as 
diffraction effects by a phenomenological boundary defined by a slit,
in each segment of the particle trajectory. This condition should be 
applied to both beam wave function and its corresponding beam
propagator $K$. The result of such a procedure is a multiple integral
that determines the actual propagator between the initial and final states
in terms of the space-time intervals due to the intermediate segments.
\bea
& & K \left( x + x_0 , T + \tau | x' , 0 \right) \nn \\
& & =
\int_{- b}^{+ b}
K \left( x + x_0 , \tau | x_0 + y_n , T + (n - 1) \tau ' \right) \nn \\
& & \quad \times K \left( x + y_n , T + (n - 1) \tau ' | \right. \nn \\
& & \quad \quad \qquad \qquad \left.
x_0 + y_{n - 1} , T + (n - 2) \tau ' \right) \nn \\
& & \quad \vdots \nn \\
& & \quad \times K \left( x + y_1 , T | x' , 0 \right) 
d y_1 d y_2 \cdots d y_n 
\label{integral}
\eea
where $\tau = n \tau '$ is the total time of revolutions $T$ is the
time necessary to insert the bunch (practically the time between two
successive bunches) and $(-b , +b)$ the space interval defining the
boundary conditions.
Obviously $b$ and $T$ are phenomenological parameters which vary from
a machine to another and must also be correlated with the geometry of the
vacuum tube where the particles circulate.

At this point we may consider two possible approximations for 
$K \left( n | n - 1 \right) \equiv 
K \left( x_0 + y_n , T + (n - 1) \tau ' | 
x_0 + y_{n - 1} + (n - 2) \tau ' \right)$:

\begin{enumerate}

\item
We substitute it with the free particle $K_0$ assuming that in the
$\tau '$ interval $(\tau ' \ll \tau)$ the motion is practically a free
particle motion between the boundaries $( -b , + b )$.

\item
We substitute it with the harmonic oscillator 
$K_{\omega} \left( n | n -1 \right)$
considering the harmonic motion of the betatronic oscillations with
frequency $\omega/{2 \pi}$

\end{enumerate}

We may notice that the convolution property~(\ref{integral}) of the
Feynman propagator allows us to substitute the multiple integral
(that becomes a functional integral for $n \longrightarrow \infty$ and 
$\tau ' \longrightarrow 0$) with the single integral
\bea
& & K \left( x + x_0 , T + \tau | x' , 0 \right) \nn \\
& & \quad =
\int_{- b}^{+ b} dy
K \left( x + x_0 , T + \tau | x_0 + y , T \right) \nn \\
& & \quad \quad \times 
K \left( x_0 + y , T | x' , 0 \right) dy
\label{single}
\eea

In this note we mainly discuss the case~1. and obtain from 
equation~(\ref{single}) after introducing the Gaussian slit
$\exp{\left[- \frac{y^2}{2 b^2}\right]}$ instead of the 
segment $\left( - b , + b \right)$ we obtain from
\bea
& & K \left( x + x_0 , T + \tau | x' , 0 \right) \nn \\
& & =
\int_{- \infty}^{+ \infty}  dy
\exp{ \left[- \frac{y^2}{2 b^2} \right]} \nn \\
& & \quad \quad \times
\left\{ \frac{2 \pi \i \hbar \tau}{m} \right\}^{- \frac{1}{2}}
\exp{\left[ \frac{\i m}{2 \hbar \tau} (x - y)^2\right]} \nn \\
& & \quad \quad \times
\left\{ \frac{2 \pi \i \hbar T}{m} \right\}^{- \frac{1}{2}}
\exp{\left[ \frac{\i m}{2 \hbar T} (x_0 + y - x')^2\right]}  \nn \\
& & = 
\sqrt{\frac{m}{2 \pi \i \hbar}}
\left(T + \tau + T \tau \frac{\i \hbar}{m b^2} \right)^{- \frac{1}{2}} \nn \\
& & \qquad \times
\exp
\left[
\frac{\i m}{ 2 \hbar} \left(v_0^2 T + \frac{x^2}{\tau} \right)
\right. \nn \\
& & \qquad \qquad \quad \quad \quad 
\left. +
\frac{\left(m^2/{2 \hbar^2 \tau^2}\right) \left(x - v_0 \tau \right)^2}
{\frac{\i m}{\hbar} \left(\frac{1}{T} + \frac{1}{\tau} \right) 
- \frac{1}{b^2}}
\right]
\label{exp}
\eea
where $v_0 = \frac{x_0 - x'}{T}$ and $x_0$is the initial central point 
of the beam at injection and can be chosen as the origin ($x_0 = 0$) of
the transverse motion of the reference trajectory in the test particle
reference frame. Where as $\hbar$ must be interpreted as the normalized 
beam emittance in the quantum-like approach. 

With an initial Gaussian profile (at $t = 0$), the beam wave function 
(normalized to 1) is
\bea
f (x) = \left\{ \frac{\alpha}{\pi} \right\}^{\frac{1}{4}}
\exp{\left[- \frac{\alpha}{2} x'^2 \right]}
\eea
r.m.s of the transverse beam and the final beam wave function is:
\bea
\phi (x) 
& = & 
\int_{- \infty}^{+ \infty} d x'
\left(\frac{\alpha}{\pi} \right)^{\frac{1}{4}}
e^{\left[- \frac{\alpha}{2} x'^2\right]}
K \left(x, T + \tau ; x', 0\right) \nn \\
& = &
B \exp{\left[C x^2 \right]}
\eea
with 
\bea
B & = &
\sqrt{\frac{m}{2 \pi \i \hbar}}
\left\{T + \tau + T \tau \frac{\i \hbar}{m b^2}\right\}^{- \frac{1}{2}}
\left\{\frac{\alpha}{\pi}\right\}^{\frac{1}{4}} \nn \\
& & \quad \times
\sqrt{
\frac{\pi}{
\left(
\frac{\alpha}{2} 
- \frac{\i m}{2 \hbar T} 
- \frac{{m^2}/{2 \hbar^2 T^2}}{
\frac{\i m}{\hbar}\left(\frac{1}{T} + \frac{1}{\tau}\right)
- \frac{1}{b^2}}
\right)
}} \nn \\
C & = &
\frac{\i m}{2 \hbar \tau}
+
\frac{{m^2}/{2 \hbar^2 T^2}}{
\frac{\i m}{\hbar}\left(\frac{1}{T} + \frac{1}{\tau}\right)
- \frac{1}{b^2}} \nn \\
& & 
+
\frac{
\frac{\tau^2}{T^2}
\left\{
\frac{{m^2}/{2 \hbar^2 T^2}}{
\frac{\i m}{\hbar}\left(\frac{1}{T} + \frac{1}{\tau}\right)
- \frac{1}{b^2}} 
\right\}^2}
{
\left(
\frac{\alpha}{2} 
- \frac{\i m}{2 \hbar T} 
- \frac{{m^2}/{2 \hbar^2 T^2}}{
\frac{\i m}{\hbar}\left(\frac{1}{T} + \frac{1}{\tau}\right)
- \frac{1}{b^2}}
\right)
}
\label{BC}
\eea

The final local distribution of the beam that undergoes the diffraction is
therefore 
\bea
\rho (x) = \left| \phi (x) \right|^2 
= B B^{*} \exp{\left[ - \tilde{\alpha} x^2 \right]}
\eea
where $\tilde{\alpha} = - (C + C^{*})$ and the total probability per 
particle is given by
\bea
P & = & \int_{- \infty} ^{+ \infty} d x \rho ( x ) 
= B B^{*} \sqrt{\frac{\pi}{\tilde{\alpha}}} \nn \\
& \approx &
\frac{1}{\sqrt{\alpha}} \frac{m b}{\hbar T}
\label{probability}
\eea
One may notice that the probability $P$ has the same order of magnitude
of the one computed in~\cite{Feynman} if $\frac{1}{\sqrt{\alpha}}$ is 
of the order of $b$.

Similarly we may consider the harmonic oscillator case 
(betatronic oscillations) compute the diffraction probability of the 
single particle from the beam wave function and evaluate the probability 
of beam losses per particle. The propagator 
$K_{\omega} \left( x + x_0 , T + \tau | x' , 0 \right)$
in the later case is:
\bea
& & K_{\omega} \left( x + x_0 , T + \tau | x' , 0 \right) \nn \\
& & \quad =
\int_{- b}^{+ b} dy
K_{\omega} \left( x + x_0 , T + \tau | x_0 + y , T \right) \nn \\
& & \qquad \quad \quad \times 
K_{\omega} \left(x_0 + y , T | x' , 0 \right) dy \nn \\
& & \quad =
\left\{
\frac{m}{2 \pi \i \hbar \omega N \sin (\omega T) \sin (\omega \tau)}
\right\}^{\frac{1}{2}} \nn \\
& & \qquad \quad \times
\exp 
\left[
\frac{\i m \omega}{2 \hbar}
\left[ 
\left(2 M - \frac{M^2}{N}\right) \left(x^2 + 2 x x_0 \right) 
\right. \right. \nn \\
& & \left. \left. \qquad \qquad \qquad \qquad \qquad \qquad \quad 
- \frac{(M - N)^2}{N} x_0^2
\right]
\right] \nn \\
& & M =
\frac{\cos (\omega \tau)}{\sin (\omega \tau)}
- \frac{1}{\sin (\omega \tau)} \nn \\
& & N =
\frac{\cos (\omega \tau)}{\sin (\omega \tau)}
- \frac{\cos (\omega T)}{\sin (\omega T)}
\label{betatron}
\eea
\section{PRELIMINARY ESTIMATES}

Preliminary numerical estimates based on the above formulae for the 
two different cases of LHC~\cite{LHC} and HIDIF~\cite{HIDIF} 
designs give the following encouraging results:

\begin{center}
{\bf LHC}
\end{center}

\begin{tabular}{lll}
Transverse Emittance, $\epsilon$ & = & $3.75$ mm mrad \\
Total Energy $E$ & = & $450$ GeV \\
T & = & $25$ nano sec.\\
b & = & $1.2$ mm \\
P & = & $3.39 \times 10^{-5}$ \\
\end{tabular}
 
\begin{center}
{\bf HIDIF}
\end{center}

\begin{tabular}{lll}
Transverse Emittance, $\epsilon$ & = & $13.5$ mm mrad  \\
Kinetic Energy $E$ & = & $5$ GeV  \\
T & = &  $100$ nano sec. \\
b & = & $1.0$ mm  \\
P & = & $2.37 \times 10^{-3}$ \\
\end{tabular}
 
\section{CONCLUSION}
These preliminary numerical results are encouraging because they
predict halo losses which seem under control. Indeed the HIDIF
scenario gives a total loss of beam power per meter which is about a
thousand higher than the LHC. However in both cases the estimated losses 
appear much smaller than the $1$ Watt/m.

\end{document}